\documentstyle[11pt]{article}
\begin{document}
\bigskip
\begin{center}
{\Large {\bf Entropy Production during Reheating at Late Times and
Neutrino Decoupling}}
\newline

{\sf Paramita Adhya $^{a,\!\!}$
\footnote{E-mail address: rinkyadh@cal2.vsnl.net.in}} and
{\sf D. Rai Chaudhuri $^{a,\!\!}$
\footnote{E-mail address: allraich@vsnl.net}}

\smallskip
$^a${\it Department of Physics, Presidency College, 86/1, College
Street,\\ Calcutta 700073, India}\\
\medskip
{\bf Abstract}
\end{center}

{\small Recent theories have proposed a variety of massive particles,
like the moduli, whose abundance or decay endangers standard
cosmological results. To dilute them, thermal inflation has been
proposed, with its own massive scalar flaton field which goes on
decaying in the era of Mev-scale temperatures. In this paper, the
effect of late-time entropy production on neutrino decoupling during
such reheating is investigated by including a term, arising from the
 rate of entropy production due
to scalar decay, in the Boltzmann equation for the
neutrino number density. 
 The effect on the
decoupling temperature of massless neutrinos is studied. It is found
that a lower bound to the scalar decay rate constant can be set at
$10^{-22}$ Gev.

}

\medskip

\begin{center}
PACS Numbers: 98.80.Cq, 98.70.Vc
\end{center}

\newpage\section{\bf Introduction} Various massive fields like the
gravitino, the Polonyi, the moduli, and the dilaton figure in
supersymmetric and string theory models\cite{gf}. The corresponding
particles are long-lived. These fields pose quite serious
cosmological problems. If they decay during baryogenesis or
nucleosynthesis(BBN), the baryon to photon ratio may be greatly
diluted in the first case and nuclear abundances distorted in the
second. If they are stable, there is a problem of over-abundance.
People have studied the problem in detail and proposed various ranges
of masses and decay rates to minimise damage to standard
cosmology\cite{GD,TB,rest}.\par An interesting proposal is that of
thermal inflation\cite{DHL1,LR}.  Quite apart from primordial
inflation\cite{riotto}, a scalar field, called the flaton, is used to
generate inflation at late times so that the potentially dangerous
fields mentioned above may be diluted away. Typically\cite{DHL}, the
inflation starts as the thermal energy density falls below the flaton
potential at the origin at a temperature of about $10^7$ Gev, and
stops when the flaton vev reaches $10^{12}-10^{14}$ Gev at a
temperature $\sim$ the flaton mass, which may be taken to be $\le
10^3$ Gev. Such a flaton field will go on decaying into the Mev-scale
temperature era, and the parameters in thermal inflation must be such
that the entropy generated due to flaton decay may not affect
nucleosynthesis.\par Recently, the effect of the decay of a massive
particle, like the flaton, on nucleosynthesis has been subjected to
detailed investigation\cite{kawa,2,k}. As the parameters and decay
schemes of such particles are yet to be generally agreed upon, it is
useful to address the general problem of the effect of late-time
entropy production, during reheating\cite{scherrer,three,four}, in
the context of BBN. In the present work, neutrino decoupling is
studied in the presence of such late-time entropy production,
 neutrino decoupling temperature being one of the key parameters 
determining BB nucleosynthesis.
Bernstein's method of ''pseudo chemical potentials''\cite{bern} is
used to introduce the entropy production rate directly into the
Boltzmann equation for the neutrino number density. The
decoupling temperature is estimated in the presence of this entropy
generation.\par The paper is organised as follows. Section 1 is this
Introduction. In Section 2, the late-time development of a universe,
once dominated by a scalar field, is set out with some modification
of the usual formalism. In Section 3, the method of ''pseudo chemical
potentials'' is applied to modify the Boltzmann equation for the
neutrino number density and incorporate into it the late-time entropy
production rate directly. In Section 4, the effect of late-time
entropy production on neutrino decoupling is studied, using
the results of the previous sections. In Section 5, conclusions are
stated.
\section {\bf Reheating at Late Times}
   In this section, a two-component universe is considered, made up
of the scalar field energy density $\rho_{\phi}$ and a radiation
energy density $\rho_R$. It is assumed \cite{kolb,DHL}, further, that
either the scalar field parameters have not supported resonance, or,
in any case, the times considered are late enough to be beyond a
possible preheating \cite{kof} interlude.\par Taking the scalar decay
rate constant to be $\Gamma$ \cite{three}, the equation for the
evolution of the scalar energy density
is\begin{equation}\frac{\partial}{\partial
t}\rho_{\phi}+3H\rho_{\phi}= - \Gamma
\rho_{\phi}.\label{eq:1}\end{equation} Defining
$\Phi=a^3\rho_{\phi}$   and   $R=a^4\rho_{\phi}$, where $a$ is the
scale factor, (\ref{eq:1}) becomes\begin{equation}\dot{\Phi}= -
\Gamma \Phi.\label{eq:2}\end{equation}

From\[ \frac{\partial}{\partial
t}[a^3(\rho_{\phi}+\rho_R)]+p_R\frac{\partial}{\partial t}a^3=0,\]
one obtains\begin{equation}
\dot{R}=a\Gamma\Phi\label{eq:3}.\end{equation}The Friedmann equation
is\begin{equation}H^2=\frac{ K^2}{a^4}(\Phi a+R),\mbox{ with }
K^2=\frac{ 8\pi}{3M_{Pl}^2}.\label{eq:4}\end{equation}\subsection{\bf
Incomplete $\Phi$-domination} In this subsection, it is assumed that
$R\ll\Phi a$, so that the square and higher powers of $ R/(\Phi a)$
can still be neglected at these times. (\ref{eq:4})
gives\[\dot{H}=-\frac{3}{2}H^2(1+\frac{\displaystyle R}{3(\Phi
a+R)}).\] With the $R\ll\Phi a$ asssumption, one can
write\begin{equation}\dot{H}\approx-\frac{3}{2}H^2(1+\frac{R}{3\Phi
a}).\label{eq:5}\end{equation}For full matter domination, at yet
earlier times, $R$ is neglected compared to $\Phi a$,
and\[\dot{H}\approx -\frac{3}{2}H^2,\]leading to $t=\frac{2}{3H}$. In
the era of incomplete $\Phi$-domination, the correction term on the
RHS of (\ref{eq:5}) is not neglected, and may be evaluated to the
approximation\begin{equation}R-R_I=\frac{dR}{da}(a-a_I),\label{eq:5a}
\end{equation}where
$t_I$ refers to some initial epoch such that $a\gg a_I$. Also, if it
is supposed that the scalar decay produces sufficiently copious
radiation, $R_I\ll R$. As a correction term is being dealt with,
these approximations should not cause much deviation from the actual
evolution. Then, for $t$ sufficiently later than $t_I$, but within
the regime under consideration, one may write, in the correction term
on the RHS of
(\ref{eq:5}),\begin{equation}R\approx\frac{\Gamma}{H}\Phi
a\label{eq:6},\end{equation}using (\ref{eq:3}) and (\ref{eq:5a}).
(\ref{eq:5})
becomes\begin{equation}\dot{H}=-\frac{3}{2}H^2(1+\frac{\Gamma}{3H}).
\label{eq:7}\end{equation}Now,
evolution will be described by the new
variable\begin{equation}x=\frac{\Gamma}{H}\label{eq:8},\end{equation}
instead of the time $t$. (\ref{eq:4}) can be written
as\begin{eqnarray}H^2 & = & \frac{K^2\Phi}{a^3}(1+\frac{R}{\Phi
a})\label{eq:8a}\\ &=&
\frac{K^2\Phi}{a^3}(1+x),\label{eq:9}\end{eqnarray} using
(\ref{eq:6}) in the correction term on the RHS of (\ref{eq:8a}).

Then, (\ref{eq:2}) leads to the evolution equation for $\Phi$\begin
{equation}\frac{d\Phi}{dx}=-\frac{2}{3+x}\Phi,\label{eq:11}\end
{equation}with the solution\begin{equation}\Phi=\Phi_I
\frac{(1+\frac{1}{3}x_I)^2}{(1+\frac{1}{3}x)^2},\label{eq:12}
\end{equation}in the approximation $x\ll 1$.
To obtain $R$, (\ref{eq:12}) is substituted in 
(\ref{eq:3}), $a$ being calculated from (\ref{eq:9}), using 
(\ref{eq:8}). The result is\begin{equation}\frac{dR}{dx}
=\frac{2}{3}(\frac{K}{\Gamma})^\frac{2}{3}x^\frac{2}{3}\Phi_I^
\frac{4}{3}\frac{(1+\frac{1}{3}x_I)^\frac{8}{3}(1+x)^{\frac{1}{3}}}
{(1+\frac{1}{3}x)^\frac{11}{3}}.\label{eq:13}\end{equation}Integration
will give $R$.
\par To compare with earlier work, at early times, one goes to the 
approximation $x,x_I\ll 1$. (\ref{eq:13}) becomes\[\frac{dR}{dx}
=\frac{2}{3}(\frac{K}{\Gamma})^\frac{2}{3}\Phi_I^\frac{4}{3}x^
\frac{2}{3}.\]On integration, taking $R_I\ll R$,
\begin{eqnarray}R & = &
\frac{2}{5}(\frac{K}{\Gamma})^\frac{2}{3}\Phi_I^\frac{4}{3}
(x^\frac{5}{3}-x_I^\frac{5}{3})\nonumber
\\ & = &
\frac{2}{5}\frac{\Gamma}{K}\Phi_I^\frac{1}{2}(a^\frac{5}{2}-a_I^
\frac{5}{2}),\nonumber\end{eqnarray}
which agrees with  \cite{kolb}.

\subsection{Incomplete Radiation domination} Now, one may consider
even later times when $\Phi a\ll R$, such that $\Phi a/R$ cannot be
neglected, but its higher powers can. The Friedmann equation
reads\begin{equation}H^2=\frac{K^2}{a^4}R(1+\frac{\Phi
a}{R})\label{eq:13'},\end{equation} which leads to
\begin{equation}\dot{H}=-2H^2[1-\frac{\Phi a}{4(\Phi
a+R)}].\label{eq:14'}  \end{equation} If, well into this epoch, the
correction term on the RHS of (\ref{eq:14'}) is neglected, the full
radiation domination relations are
found:\begin{eqnarray}H &=&\frac{1}{2t},\mbox{ and }\nonumber \\
a&=&At^{\frac{1}{2}},\label{eq:15'}\end{eqnarray}$A$ being a
constant.\par (\ref{eq:2}) has, as solution, a falling exponential
in $t$, viz. $\Phi\sim e^{-\Gamma t}$. The new evolution variable 
$x=\Gamma /H$ is now sought to be introduced in place
of $t$. To do this, instead of taking the falling exponential in $t$
directly, 
 suitable approximations to the correction
terms on the RHS of (\ref{eq:13'}) and (\ref{eq:14'}), are first
worked out. Let $t_0$ be a sufficiently late epoch,
 when $\Phi=\Phi_0\approx 0$.
Then, for use only in the correction terms, one takes
\begin{eqnarray} \Phi - \Phi_0=
\tilde{\Phi}(\frac{1}{t})-\tilde{\Phi}(\frac{1}{t_0})=
\frac{d\tilde{\Phi}}{d\frac{1}{t}}|_{t_0}(\frac{1}{t}-\frac{1}{t_0}).
\nonumber\end{eqnarray}
Neglecting $\Phi_0,1/t_0$ compared to $\Phi,1/t$, respectively,
an approximation\begin{equation}
 \Phi\approx\frac{B}{t},\label{eq:16'}\end{equation}will be used only
  in
the correction terms, i.e. in the correction terms, the falling 
exponential will be approximated by a rectangular hyperbola. B
is a constant. Using this approximation in the correction terms,
the final form of $\Phi$ will be found below to be a falling
exponential in $(1/2)x$, with a pre-exponential correction.
\par Next, a similar approximation is considered for
 $R$.
It ought to be mentioned that
$R$ refers to the total radiation present, and not only to that 
produced by decay. However, the change in $R$ is due to 
$\phi$ decay and
consequent entropy production . In the absence of this decay,
$\dot{R}=0$.\par Using (\ref{eq:15'}) and (\ref{eq:16'}) in
(\ref{eq:3}), and, integrating, one obtains, for use only in the
correction terms,\[R-R_E\approx
2AB\Gamma(t^{\frac{1}{2}}-t_E^{\frac{1}{2}}).\] If $t_E$ is
sufficiently early compared to $t$, though within the regime under
consideration, and there is sufficiently copious radiation production
since $t_E$, it is sufficient to take \[R\approx 2AB\Gamma
t^{\frac{1}{2}}\] in the correction terms. (\ref{eq:15'}) and
(\ref{eq:16'}) are now used to give, in the correction terms, 
\begin{eqnarray}\frac{R}{\Phi a}&\approx &\frac{\Gamma}{H},\mbox{ once
again, as in (\ref{eq:6}),}\nonumber \\
&=&x.\label{eq:17'}\end{eqnarray}The approximation is now $x\gg 1.$
 (\ref{eq:13'}) and (\ref{eq:14'}) become \begin{eqnarray} H^2 & = &
\frac{K^2}{a^4}R(1+\frac{1}{x}),\label{eq:10}\\
\mbox{and,  }
\dot{H} &=& -2H^2\frac{x+\frac{3}{4}}{x+1}.\nonumber\end{eqnarray}
This last equation and (\ref{eq:2})
give\begin{equation}\Phi=\Phi_E(\frac{4x_E+3}{4x+3})^{\frac{1}{8}}
e^{-\frac{1}{2}(x-x_E)}\label{eq:18'}\end{equation}One can see that
this solution is dominated by the falling exponential. The
pre-exponential factor arises because radiation domination is not
yet complete, and $(1/2)x$ is not quite $\Gamma t.$
For sufficiently late times $t,t_E$, one has $x,x_E\gg 1$,
 (\ref{eq:10}) gives $H=1/(2t)$, in the usual
way,
and (\ref{eq:18'}) reduces to \[\Phi=\Phi_Ee^{-\Gamma (t-t_E)},\]
which agrees with \cite{scherrer}.\par From
(\ref{eq:10}),\begin{equation}a  \approx (\frac{K}{H})^{\frac{1}{2}}
R^{\frac{1}{4}}(1+\frac{1}{x})^{\frac{1}{4}}.\label{eq:19'}
\end{equation}Now,
on integrating
(\ref{eq:3}), using (\ref{eq:18'}) and (\ref{eq:19'}), one has
\begin{eqnarray}R^{\frac{3}{4}}&=& R_{\infty}^
{\frac{3}{4}}\nonumber
\\
&\;&-\frac{3}{8}(\frac{K}{\Gamma})^{\frac{1}{2}}(x_E+\frac{3}{4})^
{\frac{1}{8}}\Phi_Ee^{\frac{1}{2}x_E}\int_x^{\infty}x^{\frac{3}{8}}
e^{-\frac{1}{2}x}dx,\label{eq:20'}\end{eqnarray}for
$x\gg 1$, where $R_{\infty}$ is the value of $R$ after decay is
finished. The integral in (\ref{eq:20'}) is an incomplete Gamma
function, and can be numerically evaluated for any value of $x$. \par
The temperature is defined
from\begin{equation}R=g^*\frac{\pi^2}{30}a^4T^4\label{eq:14},
\end{equation}where
$g^*$ is the effective number of relativistic degrees of freedom.\par
The entropy production originates
from \cite{early}\begin{equation}dS_{\phi}=-\frac{1}{T}d\Phi.
\label{eq:15''}
\end{equation}
\section{\bf Entropy Production and the Boltzmann Equation for the
Number Density with Scalar Decay}\par The effect of entropy production 
on decoupling is
studied here with reference to the particular problem of the electron
neutrino, $\nu,$ decoupling from a thermal bath of electrons,
positrons, and photons, in the presence of a decaying scalar field.
It is assumed that the scalar field has a very small branching ratio
into neutrinos\cite{kawa}. So, the main process which changes the
number density $n$ of the $\nu$ neutrinos is\begin{equation} \nu(k) +
\bar{\nu}(\bar{k})\longrightarrow {\cal F}(p)+\bar{\cal
F}(\bar{p}),\label{eq:16}\end{equation}where ${\cal F},\bar{\cal F}$
are fermions, and, $k$ is the energy-momentum 4-vector
$(E_k,\vec{k})$. Apart from $e^-,e^+$, the relevant fermions,
at the epoch of $\nu$ decoupling, are the
neutrinos of the other families.\par If there is no other process
like scalar decay, it is usual to assume
\cite{bern} that, in the absence of Fermi degeneracy, the
distribution functions may be written
as\begin{eqnarray}f_{\nu}(k)=f'(k) & = & e^{-\alpha '(t)-\beta
E_k}\nonumber \\ f_e(p) & = &  e^{-\beta E_p}\nonumber \\
f_{\bar{\nu}}(\bar{k})=\bar{f'}(\bar{k}) & = & e^{-\alpha '(t)-\beta
E_{\bar{k}}} \nonumber\\ f_{e^+}(\bar{p}) & = &e^{-\beta
E_{\bar{p}}},\label{eq:17}\end{eqnarray} where $T = \frac{1}{\beta}$
is the  temperature of the thermal bath, and $\alpha '$ is a
time-dependent ''pseudo chemical potential'', introduced to take into
account the departure of the decoupling neutrinos from equilibrium.
(In (\ref{eq:17}), it has been assumed that ${\cal F}$ is the
electron.)\par Decoupling is, in this case, governed by the
integrated Boltzmann
equation\begin{equation}\dot{n}+3Hn=-<\sigma|v|>(n^2-n_{EQ}^2),
\label{eq:18}\end{equation}where
$n_{EQ}$ is the equilibrium number density, 
\begin{equation}n_{EQ} = \int
\frac{gd^3k}{(2\pi)^3}e^{-\beta
E_k}=T^3/\pi^2,\label{eq:19}\end{equation}
\begin{equation}n\approx n_{EQ}e^{-\alpha
'}\cite{bern},\label{eq:23}\end{equation}and
\begin{equation}<\sigma |v|>=
\frac{1}{n_{EQ}^2}\int dKd\bar{K}e^{-\beta E_k}e^{-\beta
E_{\bar{k}}}I,\label{eq:25}\end{equation}$I$ being the invariant
integral\begin{equation}I=\int
dPd\bar{P}(2\pi)^4\delta^4(p+\bar{p}-k-\bar{k})|{\cal{M}}|^2.
\label{eq:33a}\end{equation}$g$ is the number of spin degrees of
freedom, $dK=gd^3k/[(2\pi)^32E_k]$ etc., 
and $|{\cal{M}}|^2$ the spin-averaged square of the modulus
of the relevant matrix element.\par Considering the process 
(\ref{eq:16}), and, assuming
CP-invariance and the absence of Fermi degeneracy, the Boltzmann
equation for the neutrino number density $n$ may be
written\cite{bern,early}\begin{equation}\frac{1}{a^3}
\frac{\partial}{\partial
t}(a^3n)=-\sum_{\cal F}\int dKd\bar{K}dPd\bar{P}(2\pi)^4\delta
^4(p+\bar{p}-k-\bar{k})|{\cal{M_F}}|^2(f'\bar{f}'-f\bar{f
}).\label{eq:21}\end{equation} $f'$
corresponds to the $\nu$ neutrino and $f$ to the fermion ${\cal F}$.
\\ Then,\begin{eqnarray}\frac{1}{a^3}
\frac{\partial}{\partial
t}(a^3n)&=&-\int dKd\bar{K}dPd\bar{P}(2\pi)^4\delta
^4(p+\bar{p}-k-\bar{k})|{\cal{M}}|^2(f'\bar{f}'-f\bar{f
})\nonumber\\ &-&\sum_i\int dKd\bar{K}dPd\bar{P}(2\pi)^4\delta
^4(p+\bar{p}-k-\bar{k})|{\cal{M}}_i|^2(f'\bar{f}'-f_i\bar{f
}_i)\label{eq:33b},\end{eqnarray}where, on the RHS, the 
first term refers to 
the process\begin{equation}\nu
+\bar{\nu}\longrightarrow e^-+e^+,\label{eq:16a}\end{equation}
and the second
 to the processes\begin{equation}\nu
+\bar{\nu}\longrightarrow \nu_i+\bar{\nu}_i,
\:i=\mu,\tau,\end{equation}$f_i$
 being the $\nu_i$ distribution function.
\par 
The injection of entropy will cause changes in the
distribution functions. One way to tackle 
the situation is to use the
Boltzmann equation for the $\nu$ distribution function,
 instead of the
integrated Bolzmann equation for the number density, include
$\nu+e^-(e^+)\to \nu+e^-(e^+)$ processes, and proceed numerically
\cite{kawa}. However, the integrated Boltzmann equation for the
number density will be used here to study decoupling at late times,
when the entropy generation rate is small, in the following way.
\par It has been assumed\cite{kawa}, in accordance with current 
models,
 that the $\phi$ does not decay into neutrinos. But, ref.\cite{kawa} 
finds that, for reheat temperatures $<7 Mev$, the neutrino 
distribution
function is materially affected by the decay. How can entropy injected
into the $e,e^+,\gamma$ sector affect neutrino distribution function
and, hence, neutrino density? The transfer of entropy must occur via
process (\ref{eq:16a}), and this must lead to extra terms on the RHS
of (\ref{eq:18}) if $n$ is to be affected. From (\ref{eq:21}), this
can only occur if $f$, the electron distribution function, changes
due to entropy injection.
\par The electromagnetic interactions, faster than the expansion, 
will tend quickly to thermalise the $e,e^+,\gamma$ 
(i.e. establish kinetic and chemical equilibrium in this sector).
But recent work\cite{mc,verdi} has shown that this
 thermalisation of $\phi$ decay products is not 
nearly instantaneous. A
massive $\phi$, as it goes on decaying, continually injects highly
energetic light particles, with energies much above the thermal mean.
There is no possibility of inverse decay and so the term from 
scalar decay in the Botzmann equation for, say, the electron 
distribution function opposes the equilibriating effect of the 
electromagnetic interaction.
 As long as the decay does not become negligible, complete
thermalisation of the distribution functions of the decay products
is not assured. How quickly complete thermalisation will occur, i.e.
the epoch of thermalisation, depends, not only on the relative
rapidity of the electromagnetic interactions and the expansion,
 but also on the parameters of $\phi$. 
In
fact, the requirement of thermalisation before a specific epoch (e.g.
BBN) has been used to derive bounds on these
parameters\cite{mc,verdi,k}.

\par The thermalisation of rapidly interacting, light decay products,
of a heavy boson was considered in ref.\cite{bern}. There, the
progress, in time, of the light particle distribution function $g$
was directly considered.  A solution $g=e^{-\alpha -\beta E}$ of the
Boltzmann equation was mooted, where $\alpha$ was a time-dependent
"pseudo chemical potential", although, the actual calculation was
done with a $\delta$ function distribution, because the problem was
simplified by assuming that all the heavy bosons decayed at one
instant. Following this lead, it will be supposed, 
in this paper, that
the injection of entropy into the bath, due to scalar decay, can be
taken care of by introducing a small time-dependent ''pseudo-chemical
potential'' $\alpha(t)$ into the electron and positron distribution
functions, in addition to the potential $\alpha '(t)$ already
introduced in the $\nu,\bar{\nu}$ distribution functions.
\par Of course, the electromagnetic interactions, being so fast, 
will cause $\alpha$ to be very small.\par In effect, then, the scheme
in (\ref{eq:17}) will be changed, in the absence of Fermi
degeneracy, to\begin{eqnarray}f_{\nu}(k)=f'(k)& = & e^{-\alpha '
-\beta  E_k}  \nonumber \\ f_e(p)=f(p) & = & e^{-\alpha -\beta
E_p}\nonumber \\ \bar{f'} & = & e^{-\alpha ' -\beta  E_{\bar
k}}\nonumber \\ \bar{f} & = & e^{-\alpha -\beta E_{\bar
p}}.\label{eq:29}\end{eqnarray}\par The entropy density
$s_j$, contributed by a particle $j$ with a distribution function
$h_j(q)$, may be defined, in the absence of degeneracy, 
as\[s_j=-\int g_j\frac{d^3q}{(2\pi)^3}(h_jlnh_j-h_j).\] 
Following \cite{bern},
the
covariant divergence of the entropy density current for a  component
$j$, in a process like that described by (\ref{eq:16a}), can be
shown to be\[
\frac{1}{a^3}\frac{\partial}{\partial t}(s_ja^3)\approx -\int
g_j\frac{d^3q}{(2\pi)^3}\frac{C_j}{E_j}lnh_j,\]where $C_j$ is the
relevant collision integral as defined
 $^{b,\!\!}$
\footnote{This differs from the definition in \cite{bern}
by a factor of $E_j$.}
 in \cite{early}. For example,
for the
electron,\begin{eqnarray}\frac{1}{a^3}\frac{\partial}{\partial
t}(s_ea^3) & = & -\int
dKd\bar{K}dPd\bar{P}(2\pi)^4\delta^4(p+\bar{p}-k-\bar{k})
|{\cal{M}}|^2\nonumber
\\ & \times & lnf(f'\bar{f'}-f\bar{f}).\nonumber\end{eqnarray} It can
immediately be seen that the covariant divergence of the total
entropy density current for the process (\ref{eq:16a}) is, assuming
energy
conservation,\begin{eqnarray}\frac{1}{a^3}\frac{\partial}{\partial
t}(s_{tot}a^3) & = & \int
dKd\bar{K}dPd\bar{P}(2\pi)^4\delta^4(p+\bar{p}-k-\bar{k})
|{\cal{M}}|^2\nonumber
\\ & \; & \times 2(\alpha '-\alpha) (e^{-2\alpha
}-e^{-2\alpha '})e^{-\beta
(E_k+E_{\bar{k}})}\nonumber \\ & \approx &   \int
dKd\bar{K}dPd\bar{P}(2\pi)^4\delta^4(p+\bar{p}-k-\bar{k})
|{\cal{M}}|^2\nonumber
\\ & \; & \times  2\alpha '(1-e^{-2\alpha '})
e^{-\beta (E_k+E_{\bar{k}})}\nonumber
\\ & \; &  + \int
dKd\bar{K}dPd\bar{P}(2\pi)^4\delta^4(p+\bar{p}-k-\bar{k})
|{\cal{M}}|^2\nonumber
\\ &  \; & \times (-2\alpha
)(1-e^{-2\alpha '}+2\alpha ')e^{-\beta(E_k+E_{\bar{k}})}.\label{eq:28} 
\end{eqnarray}
Here, $\alpha$, being the effect of entropy injection, is assumed to
be small at late times. Squares and higher powers of $\alpha$ have
been neglected.
\par The first term on the RHS of (\ref{eq:28}) is
the covariant divergence of the entropy density current for $\alpha
=0$, and the second term is proportional to $\alpha $. So, it can be
concluded that the second term measures, approximately, to first
order in $\alpha,$ the contribution of $\phi-$decay to the covariant
divergence of the entropy density current, corresponding to process
(\ref{eq:16a}),
viz.\begin{eqnarray}\epsilon\frac{1}{a^3}
\frac{\partial S_{\phi}}{\partial t}
 & = & -2\alpha (1-e^{-2\alpha '}+2\alpha ')\int
dKd\bar{K}dPd\bar{P}(2\pi)^4\delta^4(p+\bar{p}-k-\bar{k})
|{\cal{M}}|^2\nonumber
\\  & \; & \times
e^{-\beta(E_k+E_{\bar{k}})},\label{eq:31}\end{eqnarray}where 
$S_{\phi}$ is
defined in (\ref{eq:15''}) and $\epsilon$ is the fraction of the
entropy current from $\phi$-decay which contributes to
process(\ref{eq:16a}).\par Again, in the presence of
$\phi-$decay,\begin{eqnarray}\frac{1}{a^3}\frac{\partial}{\partial
t}(a^3n) & = & -\int
dKd\bar{K}dPd\bar{P}(2\pi)^4\delta^4(p+\bar{p}-k-\bar{k})
|{\cal{M}}|^2\nonumber
\\ & \; & \times (e^{-2\alpha '}-e^{-2\alpha })e^{-\beta
(E_k+E_{\bar{k}})}\nonumber
\\&-&\sum_i \int
dKd\bar{K}dPd\bar{P}(2\pi)^4\delta^4(p+\bar{p}-k-\bar{k})
|{\cal{M}}_i|^2(f'\bar{f}'-f_i\bar{f}_i)
\\ & = & -(n^2-n_{EQ}^2)<\sigma |v|>\nonumber\\ & -& 2\alpha\int
dKd\bar{K}dPd\bar{P}(2\pi)^4\delta^4(p+\bar{p}-k-\bar{k})|{\cal{M}}|^2
e^{-\beta(E_k+E_{\bar{k}})}\nonumber\\ &-&\sum_i \int
dKd\bar{K}dPd\bar{P}(2\pi)^4\delta^4(p+\bar{p}-k-\bar{k})
|{\cal{M}}_i|^2(f'\bar{f}'-f_i\bar{f}_i)\label{eq:32}\end{eqnarray} to
first order in $\alpha$, using (\ref{eq:23}), (\ref{eq:25}),
(\ref{eq:29}) and (\ref{eq:33b}). $<\sigma |v|>$ 
refers to (\ref{eq:16a}).\par 
Now, the finer point that, in this epoch, the
 electron neutrino has both charged and neutral current interactions,      
while the other neutrinos have only neutral current interactions
(muons and taons having already decoupled), will be neglected,
and it will be assumed that the charged and neutral current 
interactions fall out of equilibrium together. Also,
neutrinos of all types will be assumed to be massless. Then, apart
from their contributions to $\rho_R$ as separate species, the
neutrinos of the different types may be treated as identical, and it
may be supposed that they decouple together. In this approximation,
the difference between $f '$ and $f_i$ may be neglected, and the terms
within the summation sign dropped, in comparison to the other terms
on the RHS of (\ref{eq:32}).\par  So, to first order in $\alpha,$ the
integrated Boltzmann equation can be written, in the presence of
$\phi-$decay, as\begin{eqnarray}\frac{1}{a^3}\frac{\partial}{\partial
t}(a^3n) &=&
\dot{n}+3Hn\nonumber\\ &=&
-(n^2-n_{EQ}^2)<\sigma |v|> +
\frac{\epsilon}{(1-e^{-2\alpha '}+2\alpha ')}
\frac{1}{a^3}\frac{\partial S_{\phi}}{\partial
t}\label{eq:33},\end{eqnarray}from (\ref{eq:31}) and (\ref{eq:32}).
 (\ref{eq:33}) will replace (\ref{eq:18}) 
in the presence of scalar decay.
\section{\bf Effect of Scalar Decay on Neutrino Decoupling}
Decoupling will be supposed to set in when the RHS of (\ref{eq:33})
becomes less than $3Hn$. To analyse the two terms on the RHS,
radiation domination is first assumed. 
A change of variables is introduced:
\begin{equation}x=\frac{\Gamma}{H},Y=nx^{\frac{3}{2}},Y_{EQ}=n_{EQ}x
^{\frac{3}{2}}.\label{eq:39}
\end{equation}Also, (\ref{eq:10}) gives, in the era of neutrino
decoupling,
\begin{equation}H= \frac{4.461\times
10^{-19}}{Gev}T^2(1+\frac{1}{x})^{\frac{1}{2}},\label{eq:49}
\end{equation} taking $g^*=10.75$. The LHS of (\ref{eq:33}), then,
 becomes $(2\Gamma/x^{\frac{3}{2}})dY/dx$.
 Near the
start of the decoupling regime, one can put\[Y=Y_{EQ}+\Delta,\] where
$\Delta$ is very small, and it is possible \cite{early,pa1,pa2} to
set\begin{eqnarray}\frac{d\Delta}{dx} & = & 0,\mbox{ and, }\nonumber
\\ \frac{dY}{dx} & = &
\frac{dY_{EQ}}{dx}.\label{eq:47}\end{eqnarray}(\ref{eq:39}),
 (\ref{eq:49}) and (\ref{eq:19})
are used to find that $dY_{EQ}/dx>0,$ i.e. the LHS of (\ref{eq:33})
is positive. Use of (\ref{eq:23}) shows that each term on the RHS of
(\ref{eq:33}) is of the same sign. So, for the equation to hold,
\begin{equation}\alpha '>0,\end{equation}and each of the terms on the
RHS of (\ref{eq:33}) is positive. This means, that when decoupling
sets in, each term must be separately less than $3Hn$.
\par The value of $\alpha '$
chosen will determine the temperature of decoupling. This choice is
fixed as follows. In the absence of $\phi-$decay, there is only the
first term on the RHS of (\ref{eq:33}). By setting this term equal to
$3Hn$, a set of values for $\alpha '$ is obtained for neutrino
decoupling temperatures $T_D$ between $0.75$ and $3$ Mev, a range
which more or less safeguards BBN. It may be mentioned that, at
present, observed $^4He$ abundance values, $Y_p,$ fall between
$0.234$ \cite{olive} and $0.244$ \cite{izo}.  A rough estimate, as
indicated in \cite{early}, gives corresponding decoupling
temperatures $\sim 1$ Mev.\par Next, for these values of $\alpha '$,
the full RHS of (\ref{eq:33}) is put equal to $3Hn$, and the
decoupling temperatures, corresponding to different values of the
scalar decay constant $\Gamma$, are worked out.
\par The first step uses
\begin{equation}-\frac{(n^2-n_{EQ}^2)<\sigma|v|>}{3Hn}\le 1.
\label{eq:100}
\end{equation}As already discussed, in the evaluation of
$<\sigma|v|>$, it is enough to consider the process (\ref{eq:16a}),
in the s-channel with $Z$ exchange and the t-channel with $W$
exchange. Evaluating (\ref{eq:33a}), and, then, (\ref{eq:25}), using
(\ref{eq:19}),
\begin{eqnarray}<\sigma|v|> & = &
\frac{8}{\pi}G_F^2[(C_{Ve}+1)^2+(C_{Ae}+1)^2]T^2\nonumber \\ & = &
\frac{4.112 \times 10^{-10}}{Gev^4}T^2.\label{eq:44}\end{eqnarray}The
mass of the electron has been neglected. $G_F$ is the Fermi
constant.\par Taking $H=\frac{4.461\times 10^{-19}}{Gev}T^2,$ in the
absence of $\phi$-decay, and using (\ref{eq:19}), (\ref{eq:23}), and
(\ref{eq:44}), one obtains the values of $\alpha '$ shown in Table I.
\begin{table}\bf\begin{center}\begin{tabular}{|l|r|}
 \hline $T_D(Mev)$ & $\alpha '$ \\
\hline 0.75 & 4.33275\\ 
\hline  1 & 3.4705\\ 
\hline 1.25 & 2.8038 \\ 
\hline 1.5 & 2.264  \\ 
\hline  1.75 & 1.8174\\ \hline 2 & 1.4470 \\ \hline 2.25 & 1.1438\\
\hline 2.5 & 0.901\\ \hline 2.75 & 0.7109\\ \hline 3 & 0.5644 \\ 
\hline\end{tabular}\\ {\small \sf Table I: 
Neutrino $\alpha '$ values at decoupling for different decoupling 
temperatures\\ without scalar decay}\end{center}\end{table}
\par Now, the second term on the RHS of (\ref{eq:33}), i.e. the
contribution of scalar decay, is considered. Using (\ref{eq:18'}),
this term becomes\begin{eqnarray}\frac{\epsilon}{(1-e^{-2\alpha
'}+2\alpha ')}
\frac{1}{a^3}\frac{\partial S_{\phi}}{\partial
t}&=&-\frac{\epsilon}{(1-e^{-2\alpha '}+2\alpha ')}
\frac{\dot{\Phi}}{Ta^3}
\nonumber\\ &=&\frac{\epsilon}{(1-e^{-2\alpha '}+2\alpha ')}
\frac{\Gamma\Phi_E(\frac{4x_E+3}{4x+3})^{\frac{1}{8}}
e^{-\frac{1}{2}(x-x_E)}}{Ta^3}.\label{eq:101}\end{eqnarray}This term,
for $x>>1$, will be dominated by the exponential. So, rather drastic
assumptions will be made to estimate the pre-exponential, in the
absence of phenomenological information about $x_E \mbox{ and }
\Phi_E$. $x_E$ refers to any epoch sufficiently early in the regime
of incomplete radiation domination, and is set equal to 1, so that
$\Phi_E=R_E/a_E$. This is quite an approximation, because
(\ref{eq:18'}) is faithful to $\Phi\sim e^{-\Gamma t}$ provided
$x_E,x\gg 1$. With this approximation, $T_E$ is, then, estimated by
setting $x=1$ in (\ref{eq:49}). Next, one puts
$R_E=(\pi^2/30)g^*_Ea_E^4T_E^4$ and $g^*_{SE}a_E^3T_E^3=g^*_Sa^3T^3$.
This last is the weakest assumption because it is only true if there
is conservation of entropy, with $a\propto T^{-1}$ behaviour. But,
entropy is being produced, and, in fact, in the regime of $\Phi$
domination, $T\propto a^{-\frac{3}{8}}$
\cite{scherrer}. In the early part of the regime of incomplete
radiation domination, too, one expects a deviation from $a\propto
T^{-1}$. However, the expectation is that the exponential dominates
the entropy current and its effect on $\nu$ decoupling, and one can
take the view that the pre-exponential is being estimated only to
order of magnitude.  Neglect of the deviation from $a\propto T^{-1}$
behaviour in the pre-exponential, and the other approximations,
should not make too much difference to an order of magnitude
calculation of lower bounds on the scalar decay constant.\par With
all these assumptions, the condition of neutrino decoupling in the
presence of scalar decay becomes, using
(\ref{eq:49}),\begin{eqnarray}-\frac{(n^2-n_{EQ}^2)<\sigma |v|>}
{3Hn} +\frac{\epsilon}{(1-e^{-2\alpha '}+2\alpha ')}
\frac{1}{a^3}\frac{\dot{ S_{\phi}}}{3Hn}&=& \nonumber\\
\frac{208.945\: sinh\alpha '\: \Gamma_0^{\frac{3}{2}}}
{x^{\frac{1}{4}}
(x+1)^{\frac{5}{4}}}+\frac{\epsilon 10.4927x^{\frac{5}{4}}
(1+x)^{\frac{1}{4}}
e^{-\frac{1}{2}(x-1)}}{(x+\frac{3}{4})^{\frac{1}{8}}(1-e^{-2\alpha '}
+2\alpha ')
e^{-\alpha '}}&\le & 1,\label{eq:102}\end{eqnarray}
where\[\Gamma_0=\frac{\Gamma}{10^{-22}Gev}.\]It must be remembered
that $x>1$ in (\ref{eq:102}), corresponding to radiation domination.
\subsection{\bf Numerical Results}Three values of $\alpha '$ are used 
from Table I, namely, those corresponding to decoupling in  
the absence of scalar decay at $T_D=1,2,3 Mev$. In each case, the
RHS of (\ref{eq:102}) is put equal to 1, and the resulting equation
solved for $x=x_D$, taking different values of $\Gamma_0$. The
corresponding decoupling temperatures $T'_D$ are found from
(\ref{eq:49}), using $x=\Gamma/H$ to rewrite it
as\begin{equation}\frac{T}{Mev}=\frac{
14.9718\sqrt{\Gamma_0}}{(x^2+x)^{\frac{1}{4}}}.\label{eq:103}
\end{equation}
$\epsilon$ would be $1$ if the entire entropy current of decay 
contributed to process (\ref{eq:16a}). There are two points here. 
The decay modes of
$\phi$ are unknown, and it is assumed that decay products, more
massive than the electron, rapidly decouple, and almost all the
entropy current of decay ends up first in the $e,e^+,\gamma$ sector
 and,
then, goes on to contribute to the entropy current of process 
(\ref{eq:16a}) via a
non-zero $\alpha$. Also, $\alpha \ne 0$ means that there will be a
small contribution of the entropy current from $\phi-$decay to the
$e,e^+,\gamma$ processes also. But, because $\alpha$ is so small,
$\epsilon$ has been put equal to $1$.\par The
results are shown in Table II.
\begin{table} \bf \begin{center} \begin{tabular}
{|l|c|r |l|c|r |l|c|r|}  \hline 
\multicolumn{3}{|c|}{$\frac{T_D=1Mev}{\alpha '=3.4705}$} & 
\multicolumn{3}{|c|}{$\frac{T_D=2Mev}{\alpha '=1.4470}$} & 
 \multicolumn{3}{|c|}{$\frac{T_D=3Mev}{\alpha '=0.5644}$}\\ 
\hline \multicolumn{1}{|c|}
{$\Gamma_0$} & $x_D$ & $T'_D$ & \multicolumn{1}{|c|}
{$\Gamma_0$} & $x_D$ & $T'_D$ &\multicolumn{1}{|c|}
{$\Gamma_0$} & $x_D$ & $T'_D$\\ \hline 100&22414&1.00&100&5603&2.00
&100&2490&3.00\\ \hline 10&2241&1.00&10&560&2.00&10&248&3.00
\\ \hline 1&223&1.00&1&55.2&2.01
&1&24.2&3.01\\ \hline 0.1&22.6&0.985&0.25&16.3&1.83&0.5&15.4&2.65
 \\ \hline 0.02&16.5&0.513&0.1&13.7&1.26&0.1&12.9&1.30 
   
\\ \hline -&-&-&0.02&13.0&0.576&0.02&12.7&0.583
\\ 
\hline\end{tabular}\\ {\small \sf Table II:
 Decoupling Temperatures $T'_D$ in Mev for different values of\\
 the Scalar Decay parameter $\Gamma_0=\Gamma/(10^{-22}Gev)$}
 \end{center}\end{table}
The results indicate that, for values of the scalar decay constant 
$\Gamma > 10^{-22}Gev$, the neutrino decoupling temperature
is not appreciably affected by scalar decay.\par
It is necessary to check what happens if there is matter domination.
Taking $x_I<<1$ in (\ref{eq:12}), extrapolating this equation to 
$\Phi=\Phi_E$ at $x=x_E=1$, and evaluating $\Phi_E$ with the
same set of assumptions as in the case of radiation domination,
the criterion of decoupling will become, in place of (\ref{eq:102}),
\begin{eqnarray}-\frac{(n^2-n_{EQ}^2)<\sigma |v|>}
{3Hn} +\frac{\epsilon}{(1-e^{-2\alpha '}+2\alpha ')}
\frac{1}{a^3}\frac{\dot{ S_{\phi}}}{3Hn}&=& \nonumber\\
\frac{208.945\: sinh\alpha '\: \Gamma_0^{\frac{3}{2}}}{x^{\frac{1}{4}}
(x+1)^{\frac{5}{4}}}+\frac{17.3934x^{5/4}(1+x)^{1/4}}
{(1-e^{-2\alpha '}+2\alpha ')e^{-\alpha '}(1+\frac{1}{3}x)^2}\le 1.
\label{eq:104}
\end{eqnarray}Matter domination implies that $x<1$ in (\ref{eq:104}).
Writing the two terms on the RHS of (\ref{eq:104}) as A and B,
$A+B\le 1$. An examination of B will show that as the temporal
variable x increases, B increases. So, if B predominates in an epoch,
there can be no decoupling in that epoch. A little numerical work
shows that for those values of $\Gamma$ (greater than a critical
value which depends on $\alpha '$) for which A predominates over B,
A+B does not fall below 1, for values of $x<1$ (matter domination ).
So, no decoupling is possible in the epoch of matter domination.
\par The numerical work shows, therefore, that for neutrino decoupling
to proceed without significant change,
 in the presence of scalar decay, there must
 be radiation domination, in the sense that
$x=\rho_{R}/\rho_{\phi}>1$, and the scalar decay constant
$\Gamma\mbox{ must be }>10^{-22}Gev$.\par This corresponds to
reheating temperatures $>8.5 Mev$, taking for the reheating
temperature, the definition of ref.\cite{kawa} :\[
\frac{T_R}{Gev}=0.554(\frac{\Gamma}{Gev})^{\frac{1}{2}}
\sqrt{\frac{2.4\times
10^{18}}{Gev}}.\]In \cite{kawa}, it was found that the neutrino
distribution function was distorted, and
the effective number of neutrino types, $N_{eff},$ started to
decrease below three, as the reheating temperature fell below $7$
Mev, and standard BBN was endangered.  So, the present results show
broad agreement with those of
\cite{kawa} on the question of the minimum reheating temperature
which safeguards standard BBN.
\section{\bf Conclusions}The effect of
entropy injection due to scalar decay on neutrino decoupling has been
studied here by introducing  the entropy production rate term
directly into the Boltzmann equation for the neutrino {\it number
density}. The method adopted was to introduce a small ''pseudo
chemical potential'' $\alpha $ into the electron distribution
function, in addition to the standard introduction of such a
potential in the neutrino distribution function. As the
electromagnetic interactions tend to thermalise the electron
distribution function, this $\alpha $ is bound to be very small. Its
function is to transmit the entropy current, arising from scalar
decay, from the $e,e^+,\gamma$ sector, to the neutrino sector through
processes like $e^-+e^+\longrightarrow\nu +\bar{\nu}$.\par The
conclusion drawn regarding the  condition of validity of standard BBN
from this study of neutrino decoupling, viz. reheating
temperature$>8.5$ Mev, agrees broadly with the conclusions drawn in
the literature from calculation of the form of the neutrino
distribution function and the effective number of neutrino types, by
numerical integration of the Boltzmann equation for the neutrino {\it
distribution function}\cite{kawa}.\par It is found that, in the
presence of entropy injection due to scalar decay, the universe must
not be matter dominated($\rho_{\phi}>\rho_R$) in the neutrino
decoupling epoch, as in that case there is no decoupling; rather, the
universe must have attained radiation domination in the sense that
$\rho_{\phi}<\rho_R$. Consideration of the effect of scalar decay on
decoupling temperatures in the regime of radiation domination leads
to a {\it lower bound} of about $10^{-22}$ Gev on the scalar decay
constant $\Gamma.$ \bigskip \\  {\bf Acknowledgements:}\\ PA wishes
to thank Dr. S. Datta for the use of facilities at Presidency
College, Calcutta. DRC wishes to thank Rahul Biswas for help in
accessing  literature.

\end{document}